\documentclass[12pt,preprint]{aastex}

\newcommand{\kms}{\ifmmode \mathrm{km~s^{-1}}\else km~s$^{-1}$\fi}
\newcommand{\smpy}{\ifmmode M_\sun~\mathrm{yr}^{-1}\else M$_\sun$~yr$^{-1}$\fi}
\newcommand{\lir}{\ifmmode L_\mathrm{IR}\else $L_\mathrm{IR}$\fi}
\newcommand{\lsun}{\ifmmode L_\sun\else $L_\sun$\fi}
\newcommand{\msun}{\ifmmode M_\sun\else $M_\sun$\fi}
\newcommand{\nags}{\ion{Na}{1}}
\newcommand{\nad}{\ion{Na}{1}~D}

\newcommand{\cf}{\ifmmode C_f\else $C_f$\fi}
\newcommand{\co}{\ifmmode C_\Omega\else $C_\Omega$\fi}
\newcommand{\dvmax}{\ifmmode \Delta v_{max}\else $\Delta v_{max}$\fi}
\newcommand{\dvtau}{\ifmmode \Delta v_{maxN}\else $\Delta v_{maxN}$\fi}

\shorttitle{HIRES Spectroscopy of ULIRGs}
\shortauthors{Rupke \& Veilleux}

\begin{document}

\title{Keck High-Resolution Spectroscopy of Outflows in Infrared-Luminous Galaxies\footnotemark[1]}

\author{David S. Rupke and Sylvain Veilleux}
\affil{Department of Astronomy, University of Maryland, College Park, MD 20742}
\email{drupke, veilleux@astro.umd.edu}
\footnotetext[1]{The observations reported here were obtained at the W.~M. Keck Observatory, which is operated as a scientific partnership among Caltech, the University of California, and NASA. The Observatory was made possible by the generous financial support of the W.~M. Keck Foundation.}

\begin{abstract}

  Several recent studies have determined that large quantities of
  neutral gas are outflowing from the nuclei of almost all
  infrared-luminous galaxies.  These measurements show that winds in
  infrared-luminous galaxies play a significant role in the evolution
  of galaxies and the intergalactic medium at redshifts $z \ga 1$,
  when infrared-luminous galaxies dominated the star formation rate of
  the universe.  These conclusions rely on moderate resolution spectra
  ($\Delta v \geq 65$~\kms) of the \nad\ absorption line and the
  assumption that there are no unresolved, saturated velocity
  components.  For the first time, we present high resolution spectra
  ($\Delta v = 13$~\kms) of massive, infrared-luminous galaxies.  The
  five galaxies in our sample are known to host outflows on the basis
  of previous observations.  With the present observations, all \nad\
  velocity components are resolved with
  $\tau(\mbox{\nad}_1~\lambda5896) \leq 6$.  The column densities we
  measure are consistent within the errors with those measured from
  moderate-resolution observations.  This confirms that the mass,
  momentum, and energy of outflowing gas in infrared-luminous galaxies
  have been measured correctly by previous studies.

\end{abstract}


\section{INTRODUCTION} \label{intro}

Absorption-line measurements of superwinds in nearby galaxies are now
common.  Recently, large surveys have performed measurements at
moderate spectral resolution of the \nad\ doublet to study outflowing
gas in infrared-luminous galaxies
\citep{hlsa00,rvs02,m05,rvs05a,rvs05b,rvs05c}.  These surveys show
that all infrared-luminous galaxies host massive winds.  The detection
rate of blueshifted absorption lines is less than 100\% in
infrared-luminous galaxies, but this does not reflect the actual
frequency of occurrence of outflows in these galaxies, which is
$\sim$100\%.  Instead, this reflects the fact that the wind does not
cover the galaxy completely \citep{rvs05b}.  Furthermore, the mass,
momentum, and energy of the outflowing gas are large and scale with
the host galaxy's star formation rate, luminosity, and mass
\citep{rvs05b}.  Recent mid-infrared observations, as well as radio
identifications of submillimeter-selected galaxies, show that luminous
and ultraluminous infrared galaxies host most of the star formation in
the universe at $z \ga 1$ \citep{p_ea05,c_ea05}.  The prevalence of
massive winds in infrared-luminous galaxies means that these outflows
have a significant impact on the evolution of galaxies and the
intergalactic medium \citep[][and references therein]{vcb05} in the
period when most of the stars in the universe are forming.

These conclusions rely on accurate measurements of column densities
through profile fitting.  However, unresolved, narrow components can
cause an underestimate of the true column density of the absorbing gas
\citep{nh73}.  The resolution limits of these surveys are
$\geq$65~\kms\ FWHM, and the intrinsic velocity profiles are several
hundred \kms\ on average \citep{hlsa00,rvs02,m05,rvs05b,rvs05c}.
These profiles are almost certainly superpositions of many components
of smaller widths, some of which could be saturated.  A handful of
dwarf galaxies have been studied at high resolution ($10-20$~\kms;
\citealt{l_ea95,sb97,sm04}); the narrowest components in these surveys
are still resolved, with FWHM~$\ga 25$~\kms.  It is necessary to apply
the same spectral resolution to a sample of massive, infrared-luminous
galaxies, to place limits on the possible errors in column density
measurements.

For the first time, we have observed infrared-luminous galaxies at
high spectral resolution ($\Delta v = 13$~\kms).  We studied the \nad\
doublet ($\lambda\lambda5890,~5896$) in four ultraluminous infrared
galaxies (ULIRGs; $\lir/\lsun > 10^{12}$) and one luminous infrared
galaxy (LIRG; $10^{11} < \lir/\lsun < 10^{12}$) with the
high-resolution spectrograph on Keck~I.  Our primary purpose is to
find components that may be unresolved by our moderate-resolution
spectra \citep{rvs05a,rvs05b,rvs05c} and confirm or revise our column
density measurements.

\section{SAMPLE, OBSERVATIONS, AND SPECTRA} \label{samp_res}

We selected five galaxies that represent a variety of interesting
\nad\ profile types from the parent sample of
\citet{rvs05a,rvs05b,rvs05c}.  One galaxy has a near-systemic
component in \nad\ with a low-velocity blue wing.  Three show broad,
relatively smooth profiles with different numbers of components and
velocity widths.  A fifth possesses an irregular \nad\ profile with
some redshifted components.  Four have infrared luminosities dominated
by a starburst (LINER spectral type) and one is a Seyfert 2.  Their
redshifts range from 0.02 to 0.14.  These galaxies and their
properties are listed in Table \ref{sample}.

We observed them during one community-access night (27 Dec 2003 UT) at
the Keck I telescope using the High Resolution Spectrograph (HIRES;
\citealt{v_ea94}).  We used a different set-up for each galaxy, in
order to optimally align the \nad\ feature and prominent emission
lines on the detector.  The slit size was 1\farcs722 $\times$
14\arcsec\ (the D4 decker), yielding a resolution of $R \sim 23000 =
13$~\kms.  The old 2k$\times$2k CCD was in use, as well as the GG-475
filter and red collimator.  Data reduction was performed with the
MAKEE\footnote{\texttt{http://spider.ipac.caltech.edu/staff/tab/makee/index.html}}
software package.

The spectra of the \nad\ line in these five galaxies are displayed in
Figure \ref{spec}.  The moderate-resolution spectra
\citep{rvs05a,rvs05c} are plotted below.  Overlaid on these spectra
are fitted profiles, the components of each fit, and 1$\sigma$ error
bars.  The fitting procedure is described in detail in \citet{rvs05a}.
In brief, we fit multiple velocity components, assuming Gaussians in
optical depth and a constant covering fraction for each.  The fitting
is performed with the LMFIT code and measurement errors are computed
from Monte Carlo simulations.  The column densities are then computed
from the fitted parameters.  Table \ref{compprop} lists the fit
parameters and column densities of each velocity component.

Fitting only a few components ($2-4$) does a good job of matching the
profile shapes.  There are a few places in the spectra where the data
and model do not match perfectly, but these can be attributed to small
deviations from Gaussianity in the optical depth profile or a varying
covering fraction.  Power spectra of the residuals from the fit show
nothing significant.

\section{MODERATE VS. HIGH RESOLUTION}

The moderate- and high-resolution profiles are almost identical.  At a
resolution of 13~$\kms$, these galaxies have remarkably smooth \nad\
profiles.  There are some high-frequency variations in the spectra
that would appear to be marginally significant based on the 1$\sigma$
Poisson errors.  These cannot be stellar absorption lines; their
widths of tens of \kms\ or less are inconsistent with the much larger
stellar velocity dispersions of ULIRGs \citep{t_ea02}.  If they were
due to variation in the absorbing gas properties, they would not
appear in the continuum, which they do.  We conclude that these
variations are due to underestimated noise (e.g., flat-fielding
errors, fringing).

To fit the data, we increase the number of components until the
properties of one or more components becomes unconstrained in Monte
Carlo simulations \citep{rvs05a}.  Using more components causes
instabilities in the solution, and using fewer components provides an
unacceptable fit.  We find that in every case but one (F02437$+$2122)
the same number of components is required at both resolutions.  A
close examination of Figure \ref{spec} shows that corresponding
components are in general similar in velocity and shape at different
resolutions.  One difference is that high resolution gives more
sensitivity to high optical depths, which allows us to relax the
parameter boundaries imposed on the moderate-resolution observations
($\tau(\lambda 5896) \leq 5$).  Despite this, we fit only one
high-resolution component with $\tau > 5$.

A more quantitative test of the differences between high and moderate
resolution components confirms these similarities on average.  We find
that the median velocity difference per component is within a few
\kms\ of zero.  The column densities in the high-resolution data are
lower by 45\%\ on average (and do not change by more than a factor of
3).  The outflowing masses, momenta, and energies decrease by 20\%\ on
average.  These results are consistent within the errors with those
from moderate resolution spectroscopy.  We thus demonstrate that the
conclusions of recent surveys
\citep{hlsa00,rvs02,m05,rvs05a,rvs05b,rvs05c} are robust to increases
in spectral resolution.

Some noticeable differences exist between the moderate- and
high-resolution data of F02437$+$2122, including changes in profile
depth and shifts in wavelength.  These are likely due to the different
aperture sizes of the two observations: the high-resolution extraction
aperture subtends 4~arcsec$^2$, while the moderate-resolution aperture
subtends 12~arcsec$^2$.  (The differences in aperture for the other
four galaxies are negligible.)  The larger aperture may probe a
different overall velocity distribution of gas (since we observe
rotation in \nad\ in this galaxy in the moderate-resolution data) and
include more continuum light.

In one galaxy, F09039$+$0503, we resolve a narrow component that was
unresolved by our moderate-resolution spectra.  This component is
redshifted from systemic by 180~\kms, and has a (resolution-corrected)
FWHM of 32~\kms.  The FWHM from the moderate-resolution data is larger
by 50$\%$.  This component is also observed in \ion{Ca}{2}~K at
moderate resolution.  Since this component is infalling and has a
small velocity width, and given the origin of most ULIRGs in a major
merger \citep[e.g.,][]{m_ea96,c_ea96,f_ea01,b_ea02,vks02}, we suggest
that this component represents tidal debris falling back on to the
galaxy.

\section{PHYSICAL MODEL}

What motions cause the linewidths we observe?  If clouds or filaments
entrained in these outflows are dominated by internal thermal motions,
we should not expect to resolve them.  (The thermal linewidth for Na
is 2.5~\kms\ in a $T = 10^4$~K gas.)  However, the smallest velocity
FWHM we observe is a factor of 10 above this limit, and twice the
spectral resolution.  Individual velocity components must therefore be
a superposition of many clouds with different central velocities.  The
different velocities of these clouds are due to the motions of the
wind that spread the clouds over both velocity space and real space,
as seen in many simulations and observations.  Fortunately, such an
ensemble can often be properly treated as a single component as we
have done \citep{j86}.

We observe only two components that have a FWHM smaller than 100~\kms.
Dwarf starbursts have many components with FWHM $< 100$~\kms, but the
velocities of their outflows are smaller than in infrared-luminous
galaxies \citep{rvs02,sm04,m05,rvs05b}.  The larger average linewidths
in ULIRGs (FWHM $=$ 330~\kms; \citealt{rvs05b}) than in dwarf galaxies
(40~\kms; \citealt{sm04}) are a consequence of the larger energy
reservoirs available to power the outflows in infrared-luminous
galaxies \citep{rvs05b}.

A near-field example of cloud ensembles resolved into individual
components are the high-velocity clouds (HVCs) in our Galaxy.  Very
high resolution (2~\kms) \nad\ observations of intermediate- and
high-velocity clouds resolve structures with linewidths down to the
resolution limit, embedded in blended spectral structures as wide as
20~\kms\ \citep{ls_ea99}.  A width of 2~\kms\ is consistent with the
thermal linewidth of Na at $T = 10^4$~K, but implies some turbulent
broadening (or accumulation of smaller cloudlets) if the temperature
is lower than this.  Because of their proximity, HVC cloudlets are
known to be distinct not only in velocity space, but also in real
space.  \citet{ws91} find that HVCs, which have neutral hydrogen
linewidths of tens of \kms\ when observed over large angular scales,
have much smaller linewidths (a few \kms) on small angular scales
(1$\arcmin$).

Envisioning the \nad\ absorption in infrared-luminous galaxies as an
ensemble of small clouds, we can make an estimate of the number of
clouds present in the line of sight.  We compute this number by
dividing the total column density in an average ensemble (velocity
component) by the column density of an average cloud (each computed
using eq. [10] of \citealt{rvs05a}).  For a cloud, we assume an
average $\tau_{cl} = 0.1$ and $b_{cl} = 2$~\kms\ based on observations
of high-velocity clouds \citep{ls_ea99}, and for ULIRGs we measure
$\tau_{ens}(\lambda 5896) = 0.9^{+2.4}_{-0.6}$ and $b_{ens} =
200^{+170}_{-90}$~\kms\ \citep[Table 1 of][]{rvs05b}.  The average
number of clouds in a ULIRG velocity component is then 900 clouds,
with a 1$\sigma$ range of $150-4500$ clouds.  The number (230) of warm
ionized clouds or resolved structures seen in high-resolution HST
observations of NGC 3079, a well-known local galaxy hosting a
superwind \citep{cb_ea01}, falls in the low end of this range.  We
would expect our estimate for ULIRGs to be on average larger, given
that the cloud properties we assume are based on observations of very
small structures in the Galaxy.

\section{SUMMARY} \label{summary}

We have performed observations at very high spectral resolution
($\Delta v = 13$~\kms) of the \nad\ doublet in five infrared-luminous
galaxies in order to confirm or deny measurements made at moderate
resolution.  We find that the resulting column densities are
consistent within the errors with previous measurements.  Thus, the
results of previous and forthcoming studies of the \nad\ feature are
reliable \citep[e.g.,][]{hlsa00,rvs02,m05,rvs05a,rvs05b,rvs05c}.  They
find that superwinds in infrared-luminous galaxies are massive and
occur with high frequency.  The impact of these winds on galaxy
evolution and the intergalactic medium, especially at high redshifts
($z \ga 1$), is likely to be substantial.

Our observations of the \nad\ doublet do not clearly resolve any new
outflowing components in these galaxies.  One redshifted component is
newly resolved; we associate this component with tidal debris from a
merger.  These results are consistent with the lack of narrow
components (FWHM $< 25$~\kms) in nearby dwarf starbursts \citep{sm04}.
Our data are consistent with a model in which an absorber is composed
of many clouds ($\sim$1000) that fill the line of sight and form a
large ensemble that enters our fits as a single absorbing `component.'
\nad\ observations of high-velocity clouds in our Galaxy resolve
cloudlets with an average width of FWHM $= 3$~\kms, and these
cloudlets may be similar to clouds that exist in superwinds.  However,
motions that separate the clouds in velocity and real space clearly
dominate the linewidth of an individual velocity component (FWHM $=
250-350$~\kms).

DSR is supported by NSF/CAREER grant AST-9874973.  SV is grateful for
partial support of this research by a Cottrell Scholarship awarded by
the Research Corporation, NASA/LTSA grant NAG 56547, and NSF/CAREER
grant AST-9874973.  This research has made use of the NASA/IPAC
Extragalactic Database (NED), which is operated by the JPL, Caltech,
under contract with the NASA.  The authors wish to recognize and
acknowledge the very significant cultural role and reverence that the
summit of Mauna Kea has always had within the indigenous Hawaiian
community.  We are most fortunate to have the opportunity to conduct
observations from this mountain.

\clearpage

\begin{figure}
\plotone{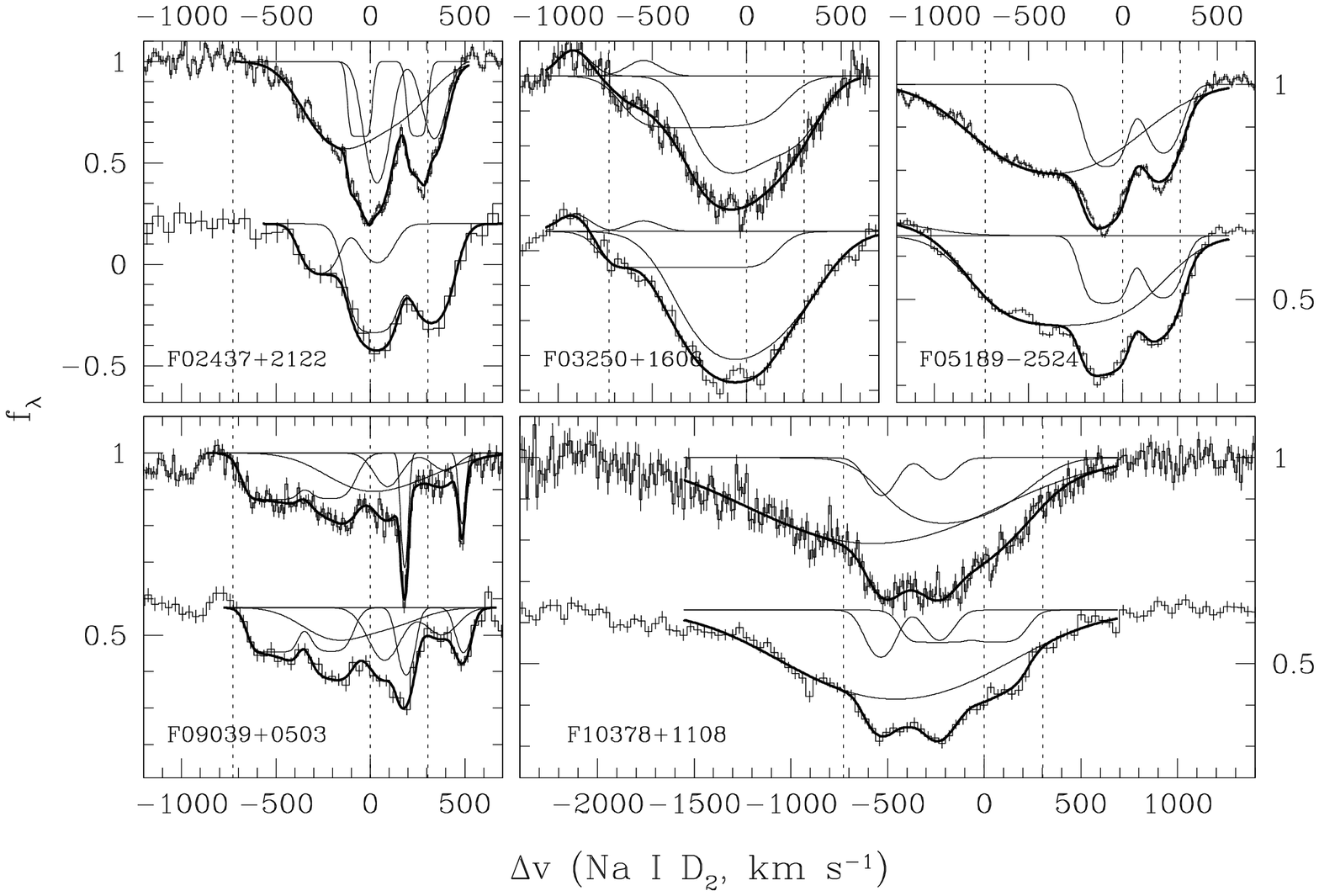}
\caption{Comparison of high- and moderate-resolution spectra.  For
  each galaxy, the high-resolution data is displayed atop the
  moderate-resolution data.  Over the spectra, we plot the fitted
  profiles, fit components, and 1$\sigma$ error bars.  The spectra are
  binned for presentation into 13~\kms\ bins (high resolution) and
  $40-60$~\kms\ bins (moderate resolution).  The moderate-resolution
  spectra also have vertical offsets applied.  The vertical dotted
  lines locate the \nad\ $\lambda\lambda5890,~5896$ doublet and
  \ion{He}{1} $\lambda5876$ emission line in the rest frame of the
  galaxy.}
\label{spec}
\end{figure}

\begin{deluxetable}{clcccrc}
\tablecaption{Sample \label{sample}}
\tablewidth{0pt}
\tablehead{
\colhead{Name} & \colhead{$z$} & \colhead{Type} & \colhead{\lir} & \colhead{$t_{exp}$} & \colhead{PA}  & \colhead{Ref} \\
\colhead{(1)}  & \colhead{(2)} & \colhead{(3)} & \colhead{(4)} & \colhead{(5)} & \colhead{(6)} & \colhead{(7)}
}
\startdata
       F02437$+$2122 &  0.0233  &       L &   11.10 &    4800 &      55 &       2 \\
       F03250$+$1606 &  0.1290  &       L &   12.13 &    9000 &       0 &       1 \\
       F05189$-$2524 &  0.04275\tablenotemark{a}& S2 & 12.11 & 3600 & 0 &       1 \\
       F09039$+$0503 &  0.1252  &       L &   12.10 &    9600 &       0 &       1 \\
       F10378$+$1108 &  0.1363  &       L &   12.32 &    6300 &       0 &       1 \\
\enddata
\tablerefs{(1) \citealt{ks98,vks99}; (2) \citealt{k_ea95,v_ea95,sm_ea03}.}
\tablecomments{Col.(1): IRAS Faint Source Catalog label.  Col.(2): Heliocentric redshift \citep{rvs05b}.  Col.(3): Optical spectral type \citep{rvs05b,rvs05c}.  L = LINER, S2 = Seyfert~2.  Col.(4): Logarithm of the infrared luminosity in units of \lsun.  Col.(5): Total exposure time, in seconds.  Col.(6): Position angle of observation slit, in degrees east of north.  Col.(7): Reference.}
\tablenotetext{a}{Our high resolution allows us to separate the \ion{Mg}{1} $b$ lines in this galaxy.  The resulting redshift is the most accurate to date ($\delta z = 0.00002$) and matches that determined from CO data ($z = 0.04275\pm0.00007$; \citealt{sss91}) and low-ionization emission lines ($z = 0.04271\pm0.00003$) within the errors.}
\end{deluxetable}

\begin{deluxetable}{lrrlrlrlrlcc}
\tabletypesize{\footnotesize}
\tablecaption{Outflow Component Properties \label{compprop}}
\tablewidth{0pt}
\tablehead{
\colhead{Name} & \colhead{$\lambda_{1,c}$} & \colhead{$\Delta v$} & \colhead{} & \colhead{$b$} & \colhead{} & \colhead{$\tau_{1,c}$} & \colhead{} & \colhead{\cf} & \colhead{} & \colhead{$N$(\nags)} & \colhead{$N$(H)}\\
\colhead{(1)} & \colhead{(2)} & \colhead{(3)} & \colhead{} & \colhead{(4)} & \colhead{} & \colhead{(5)} & \colhead{} & \colhead{(6)} & \colhead{} & \colhead{(7)} & \colhead{(8)}
}
\startdata
       F02437$+$2122 & 6032.33 &    -133 &($\pm$  4) &     222 &($\pm$  7) &    0.48 &     ($^{+0.02}_{-0.02}$) &    0.66 &($^{+0.03}_{-0.03}$) &   13.58 &   20.88 \\
             \nodata & 6034.45 &     -28 &($\pm$  1) &      39 &($\pm$  2) &    6.25 &     ($^{+1.24}_{-0.81}$) &    0.37 &($^{+0.07}_{-0.05}$) &   13.94 &   21.24 \\
             \nodata & 6036.27 &      63 &($\pm$  1) &      80 &($\pm$  3) &    0.57 &     ($^{+0.02}_{-0.02}$) &    0.88 &($^{+0.04}_{-0.04}$) &   13.21 &   20.51 \\
       F03250$+$1606 & 6650.46 &    -355 &($\pm$  5) &     224 &($\pm$ 12) &    2.50 &     ($^{+0.29}_{-0.25}$) &    0.15 &($^{+0.02}_{-0.01}$) &   14.30 &   21.61 \\
             \nodata & 6656.36 &     -89 &($\pm$  2) &     198 &($\pm$  6) &    0.93 &     ($^{+0.06}_{-0.06}$) &    0.32 &($^{+0.02}_{-0.02}$) &   13.81 &   21.13 \\
       F05189$-$2524 & 6140.41 &    -452 &($\pm$  3) &     429 &($\pm$  7) &    0.44 &     ($^{+0.02}_{-0.02}$) &    0.30 &($^{+0.01}_{-0.01}$) &   13.83 &   21.19 \\
             \nodata & 6147.86 &     -89 &($\pm$  0) &     103 &($\pm$  1) &    1.63 &     ($^{+0.05}_{-0.05}$) &    0.20 &($^{+0.01}_{-0.01}$) &   13.78 &   21.13 \\
       F09039$+$0503 & 6624.47 &    -518 &($\pm$  3) &     109 &($\pm$  8) &    4.00 &     ($^{+1.19}_{-0.79}$) &    0.13 &($^{+0.04}_{-0.03}$) &   14.19 &   21.54 \\
             \nodata & 6635.03 &     -41 &($\pm$ 12) &     282 &($\pm$ 25) &    0.10 &     ($^{+0.02}_{-0.01}$) &    0.50 &($^{+0.10}_{-0.05}$) &   13.00 &   20.35 \\
             \nodata & 6637.96 &      92 &($\pm$  4) &      98 &($\pm$ 20) &    0.05 &     ($^{+0.02}_{-0.01}$) &    1.00 &($^{+0.00}_{-0.00}$) &   12.23 &   19.58 \\
             \nodata & 6639.95 &     181 &($\pm$  0) &      22 &($\pm$  1) &    0.51 &     ($^{+0.05}_{-0.04}$) &    0.49 &($^{+0.05}_{-0.04}$) &   12.60 &   19.95 \\
       F10378$+$1108 & 6686.37 &    -673 &($\pm$ 15) &     760 &($\pm$ 40) &    0.19 &     ($^{+0.05}_{-0.02}$) &    0.48 &($^{+0.13}_{-0.04}$) &   13.72 &   21.09 \\
             \nodata & 6689.54 &    -531 &($\pm$  4) &      97 &($\pm$ 23) &    0.32 &     ($^{+0.15}_{-0.05}$) &    0.19 &($^{+0.09}_{-0.03}$) &   13.05 &   20.42 \\
             \nodata & 6695.47 &    -265 &($\pm$ 17) &     264 &($\pm$ 36) &    0.76 &     ($^{+0.15}_{-0.12}$) &    0.19 &($^{+0.04}_{-0.03}$) &   13.85 &   21.22 \\
\enddata
\tablecomments{Col.(2): Redshifted, heliocentric, vacuum wavelength of the \nad$_1$ $\lambda5896$ line, in \AA.  Col.(3): Velocity relative to systemic, in \kms.  Negative velocities are blueshifted.  Components with $\Delta v < -50$~\kms\ are assumed to be outflowing \citep{rvs05b}.  Col.(4): Doppler parameter, in \kms.  Col.(5): Central optical depth of the \nad$_1$ $\lambda5896$ line; the optical depth of the D$_2$ $\lambda5890$ line is twice this value.  Col.(6): Covering fraction of the gas.  Col.(7-8): Logarithm of column density of \nags\ and H, respectively, in cm$^{-2}$.}
\end{deluxetable}


\begin{thebibliography}{}
\bibitem[Bushouse et al.(2002)]{b_ea02} Bushouse, H.~A., et al.  2002, \apjs, 138, 1
\bibitem[Cecil et~al.(2001)]{cb_ea01} Cecil, G., Bland-Hawthorn, J., Veilleux, S., \& Filippenko, A.~V.  2001, \apj, 555, 338 
\bibitem[Chapman et al.(2005)]{c_ea05} Chapman, S.~C., Blain, A.~W., Smail, I., \& Ivison, R.~J.  2005, \apj, 622, 772 
\bibitem[Clements et al.(1996)]{c_ea96} Clements, D.~L., Sutherland, W.~J., McMahon, R.~G., \& Saunders, W.  1996, \mnras, 279, 477
\bibitem[Farrah et al.(2001)]{f_ea01} Farrah, D., et al.  2001, \mnras, 326, 1333
\bibitem[Heckman et~al.(2000)]{hlsa00} Heckman, T.~M., Lehnert, M.~D., Strickland, D.~K., \& Armus, L.  2000, \apjs, 129, 493
\bibitem[Jenkins(1986)]{j86} Jenkins, E.~B.  1986, \apj, 304, 739
\bibitem[Kim \& Sanders(1998)]{ks98} Kim, D.-C., \& Sanders, D.~B.  1998, \apjs, 119, 41
\bibitem[Kim et~al.(1995)]{k_ea95} Kim, D.-C., Sanders, D.~B., Veilleux, S., Mazzarella, J.~M., \& Soifer, B.~T.  1995, \apjs, 98, 129
\bibitem[Lehner et~al.(1999)]{ls_ea99} Lehner, N., Sembach, K.~R., Lambert, D.~L., Ryans, R.~S.~I., \& Keenan, F.~P.  1999, \aap, 352, 257 
\bibitem[Lequeux et~al.(1995)]{l_ea95} Lequeux, J., Kunth, D., Mas-Hesse, J.~M., \& Sargent, W.~L.~W. 1995, \aap, 301, 18 
\bibitem[Martin(2005)]{m05} Martin, C.~L. 2005, \apj, 621, 227
\bibitem[Murphy et al.(1996)]{m_ea96} Murphy, T.~W., Jr., Armus, L., Matthews, K., Soifer, B.~T., Mazzarella, J.~M., Shupe, D.~L., Strauss, M.~A., \& Neugebauer, G.  1996, \aj, 111, 1025
\bibitem[Nachmann \& Hobbs(1973)]{nh73} Nachmann, P., \& Hobbs, L.~M.  1973, \apj, 182, 481
\bibitem[P\'{e}rez-Gonz\'{a}lez et~al.(2005)]{p_ea05} P\'{e}rez-Gonz\'{a}lez, P.~G., et~al.  2005, \apj, 630, 82 
\bibitem[Rupke et~al.(2002)Rupke, Veilleux, \& Sanders]{rvs02} Rupke, D.~S., Veilleux, S., \& Sanders, D.~B.  2002, \apj, 570, 588
\bibitem[Rupke et~al.(2005a)Rupke, Veilleux, \& Sanders]{rvs05a} Rupke, D.~S., Veilleux, S., \& Sanders, D.~B.  2005a, \apjs, 160, 87
\bibitem[Rupke et~al.(2005b)Rupke, Veilleux, \& Sanders]{rvs05b} Rupke, D.~S., Veilleux, S., \& Sanders, D.~B.  2005b, \apjs, 160, 115
\bibitem[Rupke et~al.(2005c)Rupke, Veilleux, \& Sanders]{rvs05c} Rupke, D.~S., Veilleux, S., \& Sanders, D.~B.  2005c, \apj, in press (astro-ph/0507037)
\bibitem[Sahu \& Blades(1997)]{sb97} Sahu, M.~S., \& Blades, J.~C. 1997, \apj, 484, L125 
\bibitem[Sanders et~al.(2003)]{sm_ea03} Sanders, D.~B., Mazzarella, J.~M., Kim, D.-C., Surace, J.~A., \& Soifer, B.~T.  2003, \aj, 126, 1607
\bibitem[Sanders et~al.(1991)Sanders, Scoville, \& Soifer]{sss91} Sanders, D. B., Scoville, N. Z., \& Soifer, B. T. 1991, \apj, 370, 158 
\bibitem[Schwartz \& Martin(2004)]{sm04} Schwartz, C.~M., \& Martin, C.~L.  2004, \apj, 610, 201
\bibitem[Tacconi et~al.(2002)]{t_ea02} Tacconi, L.~J., Genzel, R., Lutz, D., Rigopoulou, D., Baker, A.~J., Iserlohe, C., \& Tecza, M.  2002, \apj, 580, 73 
\bibitem[Veilleux et~al.(2005)Veilleux, Cecil, \& Bland-Hawthorn]{vcb05} Veilleux, S., Cecil, G., \& Bland-Hawthorn, J.  2005, \araa, in press (astro-ph/0504435)
\bibitem[Veilleux et~al.(1999)Veilleux, Kim, \& Sanders]{vks99} Veilleux, S., Kim, D.-C., \& Sanders, D.~B.  1999, \apj, 522, 113
\bibitem[Veilleux et~al.(2002)Veilleux, Kim, \& Sanders]{vks02} Veilleux, S., Kim, D.-C., \& Sanders, D.~B.  2002, \apj, 143, 315
\bibitem[Veilleux et~al.(1995)]{v_ea95} Veilleux, S., Kim, D.-C., Sanders, D.~B., Mazzarella, J.~M., \& Soifer, B.~T.  1995, \apjs, 98, 171
\bibitem[Vogt et~al.(1994)]{v_ea94} Vogt, S.~S., et~al.  1994, \procspie, 2198, 362
\bibitem[Wakker \& Schwarz(1991)]{ws91} Wakker, B.~P., \& Schwarz, U.~J.  1991, \aap, 250, 484 
 \end{thebibliography}
\end{document}